\documentclass [doublespacing] {elsart}
\DeclareRobustCommand{\ion}[2]{%

\relax\ifmmode
\ifx\testbx\f@series
{\mathbf{#1\,\mathsc{#2}}}\else
{\mathrm{#1\,\mathsc{#2}}}\fi
\else\textup{#1\,{\mdseries\textsc{#2}}}%
\fi}
\begin{document}
\begin{frontmatter}

\title{Absorption spectra of Fe L-lines in Seyfert 1 galaxies}

\author[label1]{J. Dubau\corauthref{cor}},
\corauth[cor]{Corresponding author. Fax: +33-169155811}
\ead{jacques.dubau@lixam.u-psud.fr}
\author[label2]{D. Porquet}, 
\author[label3]{O. Z. Zabaydullin}

\address[label1]{LIXAM, Universit{\'e} Paris-Sud, 91405 Orsay cedex, France}
\address[label2]{SAp, CEA-Saclay, 91191 Gif-sur-Yvette, France}
\address[label3]{Institute of Nuclear Fusion, Kurchatov Institute, Moscow, 
Russia}

\begin{abstract}
Absorption L-lines of iron ions are observed, in absorption, in spectra of  
Seyfert\,1 galaxies by the new generation of X-ray satellites: Chandra (NASA) 
and XMM-Newton (ESA).  Lines associated to Fe$^{\,23+}$ to
Fe$^{\,17+}$ are well resolved. Whereas, those corresponding to Fe$^{\,16+}$ to
Fe$^{\,6+}$ are unresolved.  Forbidden transitions of the Fe$^{\,16+}$ to
Fe$^{\,6+}$ ions were previously observed, for the same objects, in the visible and infra-red 
regions, showing that the plasma had a low density. To interpret X-ray, visible 
and infra-red data, astrophysical models  assume an extended absorbing medium of 
very low density surrounding an intense X-ray source. 
We have calculated atomic data (wavelengths, radiative and autoionization rates) 
for n=2 to n'=3-4 transitions and used them to construct refined synthetic spectra of 
the unresolved part of the L-line spectra.
\end{abstract}

\end{frontmatter}

\section{Introduction}
The new generation of X-ray satellites ({\sl Chandra} and {\sl XMM-Newton})  
produces very detailed spectra, thanks to a high spectral resolution combined 
to a high sensitivity of the spectrometers on board. For the first time, it
is possible to have access to X-ray spectroscopy of extra-solar 
astrophysical objects: e.g., stellar coronae, Active Galactic
Nuclei (AGN), X-ray binaries, etc. In particular, observations of Seyfert 
galaxies, show numerous lines of highly ionized elements: predominantly in 
absorption, for Seyfert\,1 and only in emission for Seyfert\,2 galaxies.

Much atomic data have already been calculated to analyze X-ray spectra of
solar corona or laboratory plasmas. The well-known He-like ion 
diagnostics  \cite{Ga}, \cite{BD} have been extended from solar to non-solar 
coronae and also to photoionized plasmas \cite{D1}, \cite{D2}. Besides these K-lines, 
Fe-L lines have also been observed in X-ray spectra, 
for Seyfert\,1 \cite{Sa}(Fig. \ref{fig:f1}), \cite{Bl}, \cite{Ka} and for Seyfert\,2 
\cite{Ki} (Fig.\ref{fig:f2}) galaxies. To  analyze the unresolved part of the Fe-L 
spectra (Fe$^{\,16+}$ to Fe$^{\,6+}$), atomic data for the transitions from
n=2 to n'=3 have been calculated and presented as an abbreviated set, assuming a
UTA (unresolved transition array) statistical model \cite{Be}, i.e. mean wavelengths, 
statistical spectral widths of transition arrays, etc. The arguments used by the 
authors to justify statistical treatment is that various processes, such as turbulence, will 
merge lines into a broad UTA, independent of the spectral resolution of the 
measuring device.
Comparisons of the Seyfert 1 NGC 3783 spectra with different spectral resolutions
(see Tables \ref{tab:t1}), XMM \cite{Bl} and Chandra \cite{Ka} (Fig \ref{fig:f3}), 
show that the statistical assumption is not justified at least for this object. In Fig. 
\ref{fig:f3}, finer detail can be clearly resolved. Indeed, for such a low density 
plasma the number of possible absorption transitions is quite limited and as the absorption 
changes dramatically over the ionization stages (see Tables \ref{tab:t2}, 
\ref{tab:t3}) the use of a statistical width  artificially increases the real 
width of the lines. We have therefore re-calculated all the atomic data of  \cite{Be}, 
extending them to n'= 4 transitions, giving a particular importance to the 
numerous possible autoionization channels.   

\section{Seyfert Galaxies}
Seyfert galaxies, discovered by Seyfert \cite{Se}, have very compact and  bright 
centers, the so-called Active Galactic Nuclei (AGN) : their bolometric 
luminosities (i.e., including all wavelength  contributions)  L$_{\rm bol}
\sim$10$^{10}$-10$^{11}$L$_{\odot}$, L$_{\odot}$ being the solar bolometric 
luminosity. The surrounding stars are relatively faint compared to the central 
nucleus. These galaxies are relatively close to our galaxy exhibiting small red-shifts  
($z\equiv \Delta \lambda / \lambda \leq$ 0.05). Seyfert galaxy optical spectra  
show strong emission lines of ionized gas plus a non-stellar continuum.
The energy power from the galactic nucleus is commonly supposed to be due to
some super-massive black hole (M$_{\rm BH}\sim$ 10$^{6}$--10$^{9}$ M$_{\odot}$, 
(where M$_{\odot}$ is the solar mass)) surrounded by an accretion disk (see Fig. 
\ref{fig:f4}). This  disk is observed in emission in all wavelengths from X-ray 
to infra-red.

There are two types of Seyfert galaxies: the Seyfert 1 and the Seyfert 2,
distinguished by their observed visible spectra. The Seyfert 1 show both 
broad  (Full Width at Half Maximum: FWHM$\sim$2\,000---20\,000\,km\,s$^{-1}$)
and narrow (FWHM$\leq$ 2\,000\,km\,s$^{-1}$) lines, while Seyfert 2 exhibit only
narrow lines. However, Antonucci \& Miller \cite{An}, in the NGC 1068 Seyfert 2,  
detected also broad lines using polarimetry. This led to the conjoncture that 
Seyfert 1 and Seyfert 2 are the same type of objects but viewed with different angles. 
That is, for Seyfert 1, the observer views the central continuum source which is 
the ``Broad Line Region'' while, in Seyfert 2, the observer views the 
external part or the ``Narrow Line Region'', due to obscuration of the central 
continuum source  by some molecular torus (see Fig. \ref{fig:f3}). This model is 
supported by the discovery in the X-ray spectrum of a dilute and high ionization medium 
called the ``Warm Absorber'' (discovered by Halpern \cite{Ha} in MR2251-178), mainly 
observed in absorption in  Seyfert\,1,  while  only seen in emission in Seyfert\,2 (e.g., 
Kinkhabwala et al.\cite{Ki}).

\section{Calculations of atomic parameters}\label{sec:atomicdata}
\subsection{Method}\label{sec:method}
Wavelengths $\lambda$, oscillator strengths $f$ and radiative transition probabilities $A_r$
have been calculated using the SUPERSTRUCTURE code developed at UC London \cite{Ei}, 
which uses a multi-configuration expansion of the wave functions. The atomic 
Hamiltonian includes most of the Breit Pauli relativistic corrections (one-body 
and two-body terms). The non-relativistic and relativistic eigenstates 
are obtained by diagonalizing the Schr\"odinger and Breit Pauli Hamiltonian respectively. 
The matrix transformation between both eigen-states is then used to transform 
non-relativistic autoionization transition matrix elements to fine-structure 
autoionization probabilities $A_a$, in the AUTOLSJ code \cite{TFR}. The radial 
parts of the one electron wave-functions are calculated in scaled Thomas-Fermi-Dirac 
potentials, the scaling parameters, for each $l$ orbital, being derived by minimizing 
the energies of some selected LS terms.

\subsection{Results}

Calculations have been done for 11 ions from Fe$^{\,+16}$ to Fe$^{\,+6}$ using 
the ground state configuration and the excited configurations accessible by absorption 
(i.e., by electric-dipole transitions). For example:

Fe$^{\,+16}$: $1s^22s^22p^6$, $1s^22s^22p^53s$, 
$1s^22s^22p^53d$, $1s^22s2p^63p$, $1s^22s^22p^54s$, 
$1s^22s^22p^54d$, $1s^22s2p^64p$

and

Fe$^{\,+6}$:  $1s^22s^22p^63s^23p^63d^2$, $1s^22s^22p^53s^23p^63d^3$, 
$1s^22s^22p^53s^23p^63d^24s$, $1s^22s^22p^53s^23p^63d^24d$, 
$1s^22s2p^63s^23p^63d^24p$.

For Fe$^{\,+16}$, the excited configurations give bound states. Whereas for
Fe$^{\,+15}$ to Fe$^{\,+6}$, the excited configurations correspond to autoionizing
states. As examples we provide in Tables \ref{tab:t2}, \ref{tab:t3}, the wavelengths, 
absorption oscillator strengths, radiative and autoionization probabilities for 
Fe$^{\,+15}$, Fe$^{\,+14}$ and Fe$^{\,+8}$. One can observe the dramatic increase in the 
autoionization probabilities  from Fe$^{\,+15}$ to Fe$^{\,+8}$. Moreover, 
the number of possible autoionizing channels increases also dramatically. This, in 
particular, explains why one does not observe the emission lines in Seyfert\,2 
which could correspond to the absorption lines in Seyfert\,1 (see Fig. \ref{fig:f2}). 
That is, for L-lines, the photo-excited bound states, Fe$^{\,+23}$ to Fe$^{\,+16}$, decay 
by the reverse radiative transition whereas the photo-excited autoionizing states, 
Fe$^{\,+15}$ to Fe$^{\,+6}$, decay preferentially by autoionization. 

\section{Synthetic spectral modeling for the iron inner-shell 
photo-excitation lines}\label{sec:spectra}

\noindent The optical depth $\tau_{ij}(\nu)$ due to an absorption line
($i\to j$) can be written as
$$\tau_{ij}(\nu)=N_{ion}\times \sigma_{ij}(\nu)$$ where N$_{ion}$ is the ionic 
column density along the line of sight to the source (in cm$^{\,-2}$) and $\sigma_{ij}(\nu)$ 
is the photo-excitation cross-section from $i$ to $j$.

The photo-excitation (or photo-absorption) cross-section is:
\begin{equation}
\sigma_{ij}(\nu)=\frac{\pi e^{2}}{m_{e}~c}~f_{ij}~\phi(\nu)\phi(\nu), 
\end{equation}
where $e$ and $m_e$ are the electron charge and mass, 
$c$ is the speed of light, $f_{ij}$ the absorption oscillator strength, 
and $\phi(\nu)$ is a normalized line profile.

In Fig \ref{fig:f5}, the relative photo-absorption cross-sections of Fe$^{\,+16}$ to 
Fe$^{\,+6}$ are presented with the normalization factor being the same for each graph. 
If instrumental width is the dominant broadening process of the lines, 
the same gaussian profile can be used for all lines. The graph in the lower right is 
the sum of all the contributions with the assumption that each ion has the same abundance.

\section{Conclusion}

 The new generation of X-ray satellites ({\sl Chandra} and {\sl XMM-Newton}) provide us 
with higher resolution spectra than was previously available.
For the first time, we have access to high resolution X-ray spectroscopy of non-solar
objects. In particular, observations of Seyfert \,1 galaxies, show very complex spectra 
with the presence of numerous absorption lines. Accurate atomic data are crucial 
to infer most of the physical and geometrical parameters of the ``Warm Absorber''. 
We have calculated complete atomic data sets (wavelength, oscillator strength, 
auto-ionization rates) for inner-shell n=2-3 and n=2-4 (mainly 2p--3d, 2p--4d) 
photo-excitation for Fe ions (from 10 electrons to 20 electrons). Observations in 
UV, visible and infra-red wavelengths, where those Fe ions emit the most, are 
also important to have a realistic plasma modeling.

\clearpage

\begin{table}
\caption{Energy (keV) and wavelength (\AA) ranges, as well as the 
spectral resolution of the spectrometers aboard the new generation of X-ray satellites:
{\sl Chandra} (NASA) and {\sl XMM-Newton} (ESA). 
The {\sl LETG} and the {\sl HETG} are onboard {\sl Chandra},
 and the {\sl RGS} is on board {\sl XMM-Newton}.}
\label{tab:t1}
\vspace*{1cm}
\begin{center}
\begin{tabular}{cccc}
\hline
                       &{\sl LETG}&{\sl HETG}& {\sl RGS} \\
E (keV)                & [0.07 -- 8.86] &  [0.4 -- 10.0] & [0.35 -- 2.5]         \\
$\lambda$ (\AA)        & [1.4  -- 170]  &  [1.2 -- 31]   & [5 -- 38]        \\
$\Delta \lambda$ (\AA) &  0.05         & 0.012-0.023  &  0.06       \\ 
\hline  
\end{tabular}
\end{center}
\end{table}
\begin{table}
\label{}
\caption{Absorption oscillator strengths, wavelengths, radiative and autoionization 
probabilities}
\label{tab:t2}
\begin{center}
\begin{tabular}{ccccc}
\hline
(from the ground level of Fe$^{\,15+}$)& $1s^22s^22p^63s  \,\,^2S_{1/2}$\cr
to the upper level & f (abs) & A$_{\rm r}$(s$^{-1}$)&$\lambda$(\AA)& A$_{\rm a}$(s$^{-1}$)\cr                   
\hline
$1s^22s^22p^53s^2\,\,^2P_{3/2}$&      0.07&    7.96 (+11)&     17.20& 8.50 (+12)\cr
$1s^22s^22p^53s^2\,\,^2P_{1/2}$&      0.04&    8.76 (+11)&     16.92& 8.71 (+12)\cr
$1s^22s^22p^53s3d\,\,^2P_{3/2}$&      0.06&    8.74 (+11)&     15.44& 1.95 (+12)\cr
$1s^22s^22p^53s3d\,\,^4D_{1/2}$&      0.12&    3.47 (+12)&     15.36& 7.46 (+05)\cr
$1s^22s^22p^53s3d\,\,^4D_{3/2}$&      0.22&    3.08 (+12)&     15.34& 4.04 (+10)\cr
$1s^22s^22p^53s3d\,\,^2D_{3/2}$&      0.35&    5.03 (+12)&     15.20& 2.13 (+12)\cr
$1s^22s^22p^53s3d\,\,^2P_{1/2}$&      0.94&    2.75 (+13)&     15.09& 4.40 (+12)\cr
$1s^22s^22p^53s3d\,\,^2P_{3/2}$&      1.50&    2.20 (+13)&     15.07& 1.82 (+12)\cr
$1s^22s2p^63s3p  \,\,^2P_{1/2}$&      0.08&    2.87 (+12)&     13.96& 5.87 (+13)\cr
$1s^22s2p^63s3p  \,\,^2P_{3/2}$&      0.19&    3.20 (+12)&     13.94& 4.10 (+13)\cr
$1s^22s^22p^53s4d\,\,^4D_{1/2}$&      0.12&    5.11 (+12)&     12.48& 1.32 (+12)\cr
$1s^22s^22p^53s4d\,\,^2P_{3/2}$&      0.25&    5.31 (+12)&     12.46& 1.20 (+12)\cr
$1s^22s^22p^53s4d\,\,^4D_{3/2}$&      0.29&    6.38 (+12)&     12.33& 1.49 (+12)\cr
$1s^22s^22p^53s4d\,\,^2P_{1/2}$&      0.16&    7.22 (+12)&     12.33& 1.51 (+12)\cr
$1s^22s2p^63s4p  \,\,^2P_{3/2}$&      0.05&    1.35 (+12)&     11.19& 5.60 (+11)\cr
$1s^22s2p^63s4p  \,\,^2P_{1/2}$&      0.03&    1.44 (+12)&     11.19& 5.02 (+11)\cr
\hline
\end{tabular}
\end{center}
\end{table}
\begin{table}
\caption{Absorption oscillator strengths, wavelengths, radiative and autoionization 
probabilities}\label{tab:t3}
\begin{center}
\begin{tabular}{ccccc}
\hline
(from the ground level of Fe$^{\,14+}$)& $ 1s^22s^22p^63s^2 \,\,^1S_{0}$\cr
to the upper level & f (abs) & A$_{\rm r}$(s$^{-1}$)&$\lambda$(\AA)& A$_{\rm a}$(s$^{-1}$)\cr                   
\hline
$1s^22s^22p^53s^23d  \,\,^3D_{1}$&      0.59&    5.49 (+12)&     15.51& 7.75 (+12)\cr
$1s^22s^22p^53s^23d  \,\,^1P_{1}$&      2.55&    2.44 (+13)&     15.26& 1.43 (+13)\cr
$1s^22s2p^63s^23p    \,\,^1P_{1}$&      0.27&    3.05 (+12)&     14.09& 8.99 (+13)\cr
$1s^22s^22p^53s^24d  \,\,^3D_{1}$&      0.37&    5.12 (+12)&     12.74& 8.49 (+12)\cr
$1s^22s^22p^53s^24d  \,\,^1P_{1}$&      0.41&    5.80 (+12)&     12.60& 8.78 (+12)\cr
$1s^22s2p^63s^24p    \,\,^1P_{1}$&      0.08&    1.42 (+12)&     11.38& 8.82 (+13)\cr
\hline
(from the ground of Fe$^{\,8+}$) &$1s^22s^22p^63s^23p^6 \,\,^1S_{0}$\cr
to the upper level & f (abs) & A$_{\rm r}$(s$^{-1}$)&$\lambda$(\AA)& A$_{\rm a}$(s$^{-1}$)\cr                   
\hline
$1s^22s^22p^53s^23p^63d  \,\,^3D_{1}$&      0.66&    5.37 (+12)&     16.59& 4.19 (+14)\cr
$1s^22s^22p^53s^23p^63d  \,\,^1P_{1}$&      1.52&    1.27 (+13)&     16.33& 4.77 (+14)\cr
$1s^22s^22p^53s^23p^64d  \,\,^1P_{1}$&      0.26&    2.67 (+12)&     14.59& 4.05 (+14)\cr
$1s^22s^22p^53s^23p^64d  \,\,^3D_{1}$&      0.18&    1.89 (+12)&     14.39& 4.00 (+14)\cr
$1s^22s2p^63s^23p^64p    \,\,^1P_{1}$&      0.06&    8.52 (+11)&     12.80& 6.20 (+13)\cr
\hline
\end{tabular}
\end{center}
\end{table}

\clearpage

\begin{figure*}
\caption{The {\sl RGS} ({\sl XMM-Newton}) first order spectrum of Seyfert 1 IRAS 13349+2438 
corrected for  cosmological redshift (z = 0.10764), from Sako et al. \cite{Sa}. The 
wavelength bins are approximately 0.1 {\AA} wide. A best-fit model spectrum is 
superimposed in red.}
\label{fig:f1}
\end{figure*}\begin{figure*}
\caption{Effective-area-corrected, first-order ({\sl XMM/RGS~1} (red) and ({\sl XMM/RGS~2})
(blue)) first-order spectra of Seyfert 2 NGC~1068 shifted to its rest frame ($z=0.00379$), from
Kinkhabwala et al. \cite{Ki}. The spectral discontinuities are due to chip gaps in the 
CCD arrays, bad pixels, etc.}
\label{fig:f2}
\end{figure*}
\begin{figure*}
\caption{{\sl Chandra/HETG} spectrum of Seyfert 1 NGC 3783 binned to 0.01 {\AA} focusing 
in the wavelength range [15-18\,\AA] from Kaspi et al. \cite{Ka}. Broad absorption 
feature from blended inner-shell 2p3d absorption lines of Fe ions.}

\label{fig:f3}
\end{figure*}
\begin{figure*}
\caption{A schematic diagram of the Unified Scheme of AGN, from
Urry \& Padovani \cite{Ur}. Surrounding the central supermassif black hole is a 
luminous accretion disk. The so-called ``Broad Line Region'' and ``Narrow Line Region''
produces broad lines and narrow lines observed in the optical range. 
The contour of the ``Warm Absorber'' observed in X-ray is also shown, as well as,
jets only seen in Radio-Loud objects.}
\label{fig:f4}
\end{figure*}
\begin{figure*}
\caption{Photo-absorption cross-sections of the differents Fe ions (Fe$^{\,+16}$ to 
Fe$^{\,+6}$) constituing the unresolved Fe-L feature, versus wavelengths. The final graph 
is built from the former graphs, assuming the same abundance for each ion.}
\label{fig:f5}
\end{figure*}

\vfill

\clearpage

\vspace{3cm}
\begin{figure*}[thbp]
\begin{center}
\begin{picture}(200,210)
\put(0,0){\includegraphics{Fig1.ps}}
\put(-180,-250){\framebox(560,400)}
\end{picture} 
\end{center}
\vspace{9cm}
\begin{center}
{\Large \bf Figure 1}
\end{center}
\end{figure*}

\begin{figure*}[thbp]
\begin{center}
\begin{picture}(200,210)
\put(0,0){\includegraphics{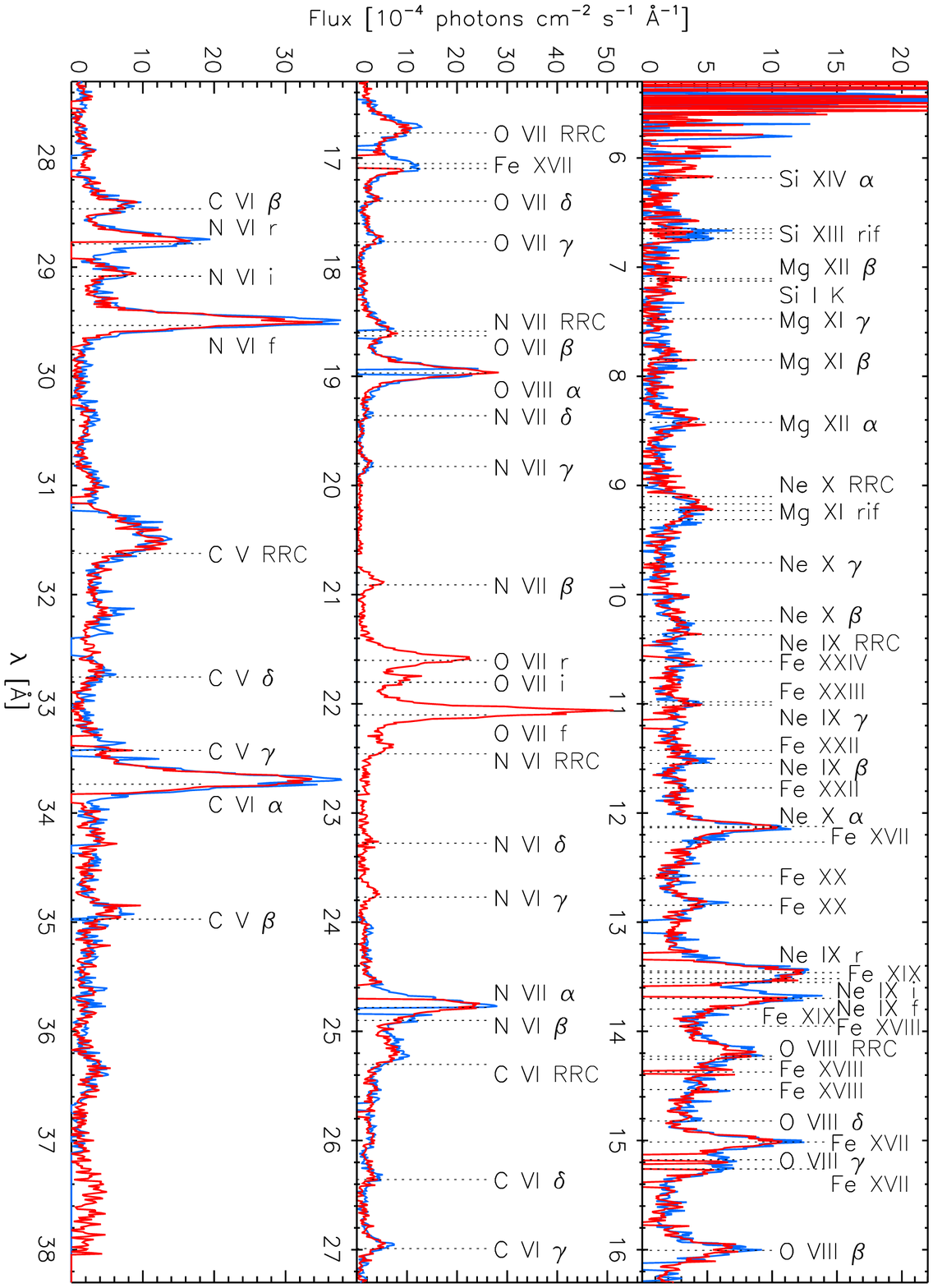}}
\put(-180,-69){\framebox(560,420)}
\end{picture} 
\end{center}
\vspace{3cm}
\begin{center}
{\Large \bf Figure 2}
\end{center}
\end{figure*}

\begin{figure*}[thbp]
\begin{center}
\begin{picture}(200,210)
\put(0,0){\includegraphics{Fig3.ps}}
\put(-150,-69){\framebox(500,400)}
\end{picture} 
\end{center}
\vspace{3cm}
\begin{center}
{\Large \bf Figure 3}
\end{center}
\end{figure*}

\begin{figure*}[thbp]
\begin{center}
\begin{picture}(200,210)
\put(0,0){\includegraphics{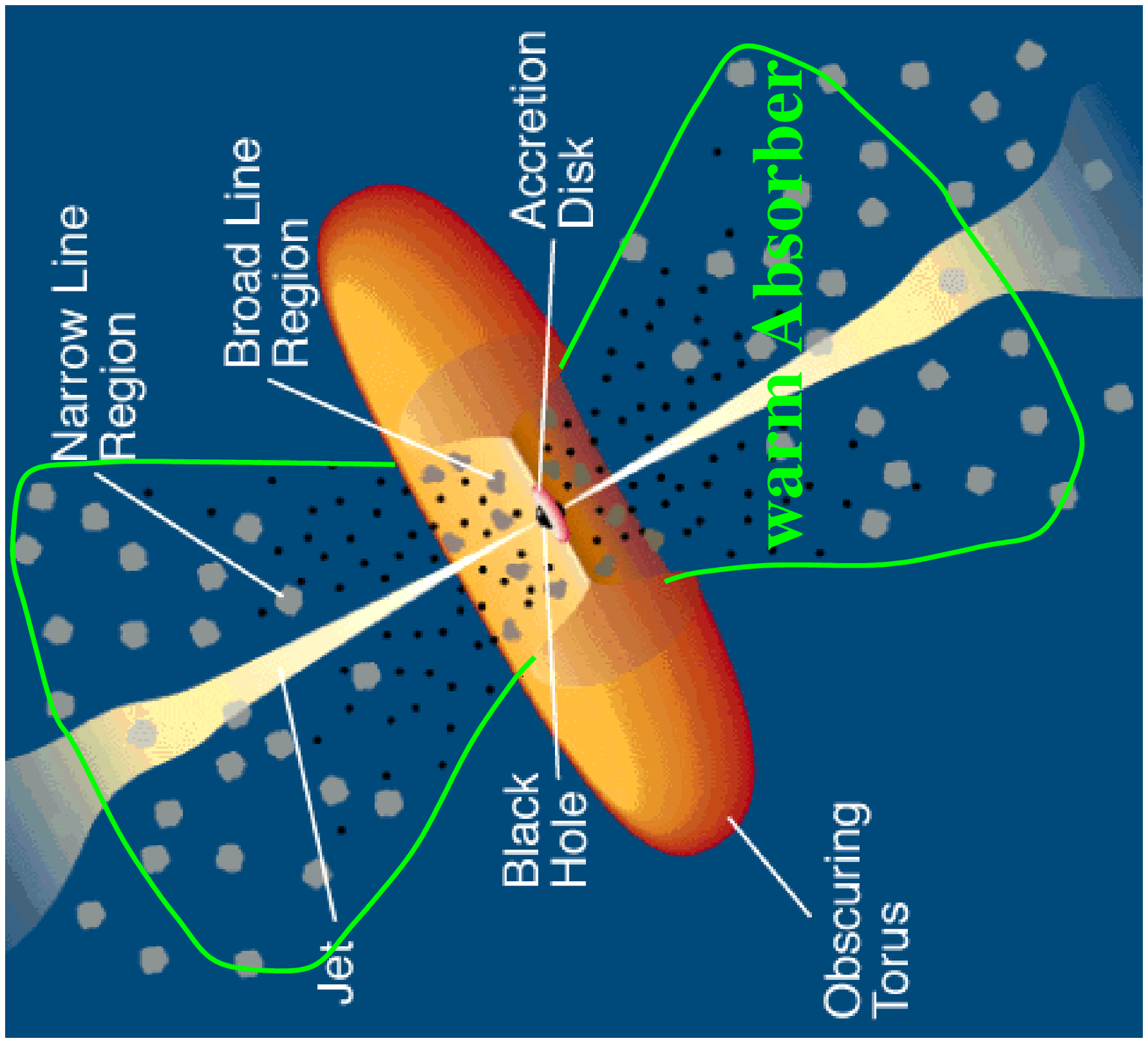}}
\put(-150,-100){\framebox(470 ,460)}
\end{picture} 
\end{center}
\vspace{4cm}
\begin{center}
{\Large \bf Figure 4}
\end{center}
\end{figure*}

\begin{figure*}[thbp]
\begin{center}
\begin{picture}(200,210)
\put(10,0){\includegraphics{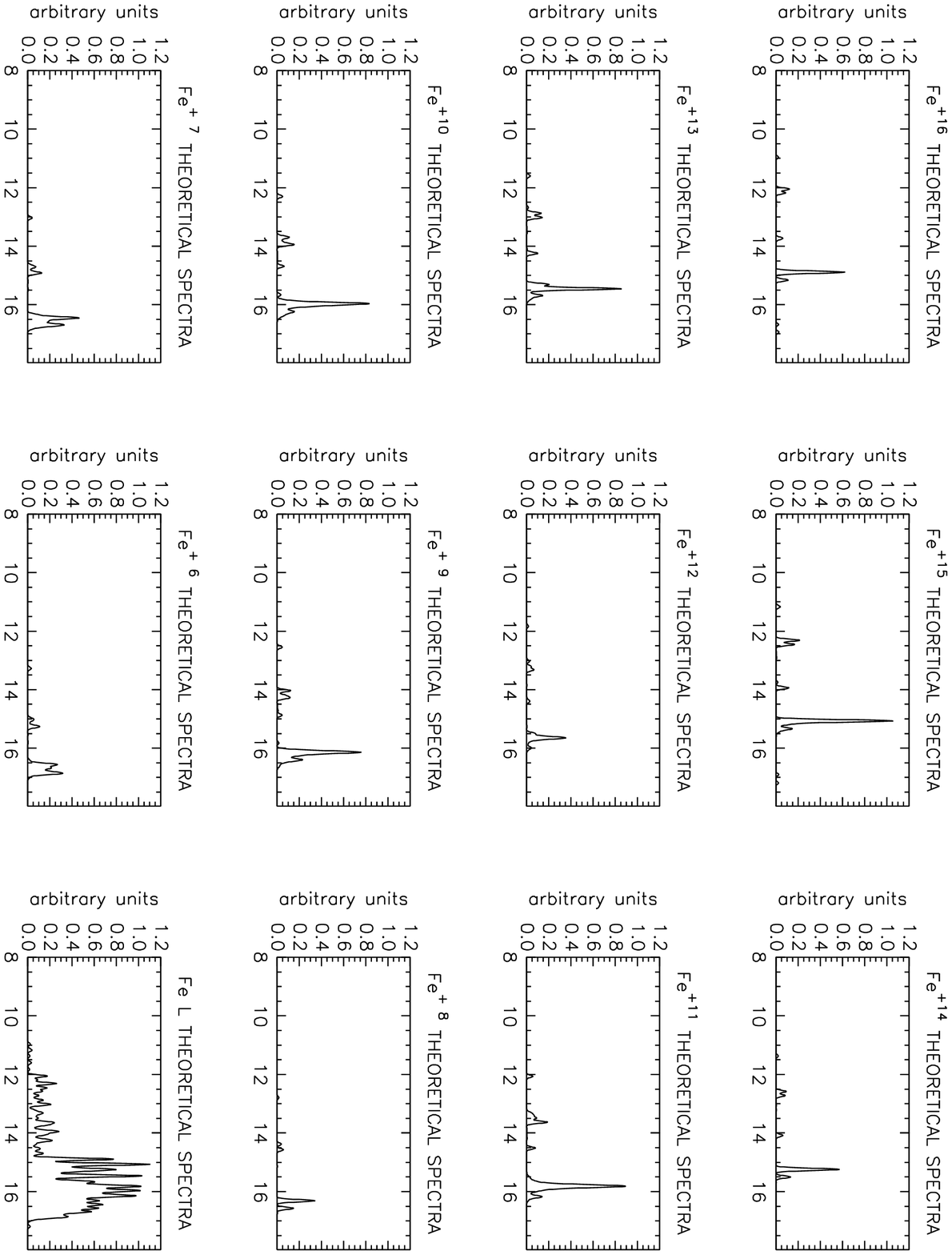}}
\put(-170,-69){\framebox(560,420)}
\end{picture} 
\end{center}
\vspace{3cm}
\begin{center}
{\Large \bf Figure 5}
\end{center}
\end{figure*}

\end{document}